\documentclass[twocolumn,showpacs,preprintnumbers,amsmath,amssymb]{revtex4}

\usepackage{graphicx}
\usepackage{dcolumn}
\usepackage{bm}
\usepackage{amssymb}
\usepackage{amssymb}

\def\>{\rangle}

\newcommand{\ucite}[1]{$^{\mbox{\scriptsize \cite{#1}}}$}
\newtheorem{theorem}{Theorem}
\newtheorem{lemma}[theorem]{Lemma}

\newtheorem{definition}{Definition}

\newtheorem{remark}{Remark}

\begin{document}

\title{Quasi multipartite entanglement measure based on quadratic functions}

\author{Jing Zhang}
 \email{zhangjing97@mails.tsinghua.edu.cn}
 \affiliation{%
Department of Automation, Tsinghua University, Beijing 100084, P.
R. China
}

\author{Chun-Wen Li}%
\affiliation{%
Department of Automation, Tsinghua University, Beijing 100084, P.
R. China
}%

\author{Re-Bing Wu}%
\affiliation{%
Department of Chemistry, Princeton University, Princeton, NJ
08544, USA
}%

\author{Tzyh-Jong Tarn}%
\affiliation{%
Department of Electrical and Systems Engineering, Washington
University, St. Louis, MO 63130, USA
}%

\author{Jian-Wu Wu}%
\affiliation{%
Department of Automation, Tsinghua University, Beijing 100084, P.
R. China
}%

\date{\today}

\begin{abstract}
We develop a new entanglement measure by extending Jaeger's
Minkowskian norm entanglement measure. This measure can be applied
to a much wider class of multipartite mixed states, although still
"quasi" in the sense that it is still incapable of dividing
precisely the sets of all separable and entangled states. As a
quadratic scalar function of the system density matrix, the quasi
measure can be easily expressed in terms of the so-called
coherence vector of the system density matrix, by which we show
the basic properties of the quasi measure including (1)
zero-entanglement for all separable states, (2) invariance under
local unitary operations, and (3) non-increasing under local POVM
(positive operator-valued measure) measurements. These results
open up perspectives in further studies of dynamical problems in
open systems, especially the dynamic evolution of entanglement,
and the entanglement preservation against the environment-induced
decoherence effects.
\end{abstract}

\maketitle

\section{Introduction}
\label{s1}

In recent years, research on quantum information science\ucite{r1}
has made a tremendous leap inspired by its potential impacts on
the existing information technologies. Among the rich theoretical
studies from various fields, quantum
entanglement\ucite{r2,r3,r4,r5,r6,r7,r8,r9,r10}, as a pure quantum
phenomenon, has been recognized to be the key of high computation
ability and communication security in quantum information
implementations. It is practically very important to quantify the
amount of entanglement in quantum states that embodies the
capacity of quantum information process. Considerable efforts have
been addressed in the literature. For two-qubit pure states, this
problem has been completely solved in various ways such as the
partial entropy entanglement measure\ucite{r11}. But for more
general quantum states, i.e. mixed and/or multipartite states, the
solution is far from perfect. The difficulties are mainly due to
the fact that the set of the separable multipartite mixed states
is too erose to be distinguished from that of the entangled
states.

The existing definitions of the entanglement measures can be
roughly classified into three classes. The first class comes from
the physical intuitions. For examples, the entanglement
cost\ucite{r11,r12} is defined as the average number of EPR pairs
required to approximate the system state $\rho$; the distillable
entanglement\ucite{r11,r13,r14} is defined as the average number
of approximate EPR pairs obtainable from $\rho$. The second class
appoints the extremum of some minimization problem as the measure,
e.g. the entanglement of formation\ucite{r11,r15} and the relative
entropy of entanglement\ucite{r16,r17,r18,r19,r20}. The third
class is related to the eigenvalues of some matrix, e.g.
concurrence\ucite{r15} and the entanglement measures based on the
PPT(positive partial transpose) condition\ucite{r21,r22,r23}.

The above approaches are physically intuitive, but not convenient
in calculations. Following the idea of Wootters\ucite{r15},
Jaeger\ucite{r24,r25,r26} proposed a novel entanglement measure
for multi-qubit states. This measure can be explicitly expressed
in terms of the Stokes parameters or, equivalently, the so-called
coherence vector\ucite{r27,r28,r29,r30,r31} of the system density
matrix. The measure is well-defined for multi-qubit pure states,
however, not directly applicable to mixed states.

In this paper,we are going to develop an entanglement measure that
fulfils certain conditions for general quantum states. Our measure
is extended from Jaeger's measure, which can also be explicitly
expressed as a quadratic function of the coherence vector of
quantum states under consideration. Since this measure is not yet
perfect for some entangled states, even if it is better than
Jaeger's measure, we prefer to call it quasi entanglement measure.
The paper is organized as follows: in section \ref{s2}, the
concept of quasi entanglement measure is introduced. In section
\ref{s3}, we present the (expanded) coherence vector
representation for multipartite density matrices, by which the
quasi entanglement measure introduced in section \ref{s2} is
re-expressed. In this picture, properties of this measure are
discussed in section \ref{s4} followed by several typical examples
in section \ref{s5}. Finally, summary and perspectives for future
studies are given in section \ref{s6}.

\section{Quasi entanglement measure}
\label{s2}

Basically, in quantum physics, separable states refer to quantum
states that can be prepared by classical means such as local
unitary operations and local measurements. In the mathematical
language, a $n$-partite separable state $\rho$ can be written
as\ucite{r32}:
\begin{equation}\label{Separable states}
\rho=\sum_i p_i
|\psi_i^1\rangle\langle\psi_i^1|\otimes\cdots\otimes|\psi_i^n\rangle\langle\psi_i^n|,\,\,\,\sum_i
p_i=1,
\end{equation}
where $p_i\geq 0$ and $|\psi_i^k\rangle$ is a pure state of the
$k^{th}$ subsystem. The quantum states that are not separable are
called entangled states. For a general entangled state $\rho$, a
perfect entanglement measure $E(\rho)$ should satisfy the
following conditions\ucite{r16}:
\begin{itemize}
    \item $E(\rho)\geq 0$; $E(\rho)=0$ if and only if $\rho$ is separable;
    \item Local unitary operations leave $E(\rho)$
    invariant, i.e. $E(\rho)=E(U
 \rho
    U^{\dagger})$ for arbitrary $U=U_1\otimes\cdots\otimes
    U_n$, where $U_i$ is a unitary transformation acting on the
    $i^{th}$ subsystem;
    \item $E(\rho)$ is non-increasing under LOCC (local operation and classical
    communication) operations. Note that the LOCC operation can be mathematically expressed as $\Theta(\rho)=\sum_r M_r \rho
    M_r^{\dagger}$, where $M_r=L_{r,1}\otimes\cdots\otimes L_{r,k}$ and $\sum_{r}
 M_r^{\dagger} M_r=I$.
\end{itemize}

Among the enormous efforts to seek entanglement measures that
fulfil the above criteria, Jaeger\ucite{r24,r25,r26} proposed a
scheme to measure entanglement in multi-qubit states borrowing
ideas from Wootters\ucite{r15}:
$$E(\rho)=tr\rho F(\rho),$$
where $\quad F(\rho)=\sigma_y^{\otimes n}\rho^*\sigma_y^{\otimes
n}$ is the flip operation on $\rho$ in which $\sigma_y$ is the
$y$-axis Pauli matrix and $\rho^*$ denotes the complex-conjugate
of the system density matrix $\rho$.

Jaeger's measure is a good measure for pure multi-qubit states,
and, in particular, coincides with the so-called concurrence
squared\ucite{r15} for pure two-qubit states. The remarkable
advantage is that the measure can be expressed as the Minkowskian
norm of the Stokes parameters or, equivalently, the coherence
vector of $\rho$, which are easy to be computed in practice.

However, Jaeger's measure fails to precisely quantify entanglement
in general mixed states, because it might be non-zero for
separable mixed states. For example, one can verify that his
measure is equal to $\frac{1}{2^n} \neq 0$ for the separable and
completely mixed state $\frac{1}{2^n}I_{2}^{\otimes n}$ where
$I_{2}$ is the two-dimensional identity matrix.

Actually, having an insight into the definition, we may find that
the flip operation $F(\rho)$ for multi-qubit pure states flips a
separable pure state $$|\psi\rangle\langle\psi |=\otimes_{k=1}^n
|\psi^k\rangle\langle\psi^k|,$$
to the separable pure state
$$|\widetilde{\psi}\rangle\langle\widetilde{\psi} |=\otimes_{k=1}^n
|\widetilde{\psi}^k\rangle\langle\widetilde{\psi}^k|,$$
where
$|\psi^k\rangle$ and
$|\widetilde{\psi}^k\rangle=\sigma_y(|\psi^k\rangle)^*$ satisfy:
$$\langle
\psi^k|\widetilde{\psi}^k \rangle=0, \quad
|\psi^k\rangle\langle\psi^k|+|\widetilde{\psi}^k\rangle\langle\widetilde{\psi}^k|=I.$$
Hence, for pure separable states, we have:
$$E(|\psi\rangle\langle\psi|)=|\langle\psi|\widetilde{\psi}\rangle|^2=\prod_{k=1}^n |\langle
\psi^k|\widetilde{\psi}^k \rangle|^2=0.$$ However, the measure
becomes much more complicated for mixed separable states. In fact,
for mixed separable states in the form of (\ref{Separable
states}), one can find that:
$$E(\rho)=\sum_{i,j}p_i p_j \prod_k|\langle \psi_i^k | \widetilde{\psi}_j^k\rangle|^2,$$
in which $\langle\psi^k_i|\widetilde{\psi}^k_j\rangle$ is
generally non-zero for $i\neq j$. Hence the measure $E(\rho)$ is
improper because it usually gives positive values for mixed
separable states.

The main reason for the above imperfection is that there are more
than one term in the decomposition (\ref{Separable states}). This
reflects the mixedness of the state that is somewhat related to
the classical correlation information\ucite{r16} and should not be
taken into account for the measure of quantum entanglement. In
this regard, we will remove this amount of information from the
original Jaeger's measure in order to obtain a better one for more
general multipartite mixed states.

Firstly, we generalize the aforementioned "flip" operation to
multipartite systems:
\begin{definition}
The "flip" operation $F(\rho)$ and the "unflip" operation
$\bar{F}(\rho)$ on the multipartite density matrix $\rho$ are
expressed as follows:
\begin{eqnarray}\label{Flip operation}
F(\rho)=\sum_{\begin{array}{c}
  \scriptscriptstyle 1\leq i_1<j_1 \leq N_1 \\
  \scriptscriptstyle \cdots \\
  \scriptscriptstyle 1\leq i_n<j_n \leq N_n \\
\end{array}}\left(\bigotimes_{k=1}^n\sigma_{i_k
j_k}^{(k)}\right)\rho^{*}\left(\bigotimes_{k=1}^n\sigma_{i_k
j_k}^{(k)}\right), \\
\bar{F}(\rho)=\sum_{\begin{array}{c}
  \scriptscriptstyle 1\leq i_1<j_1 \leq N_1 \\
  \scriptscriptstyle \cdots \\
  \scriptscriptstyle 1\leq i_n<j_n \leq N_n \\
\end{array}}\left(\bigotimes_{k=1}^n\bar{\sigma}_{i_k
j_k}^{(k)}\right)\rho\left(\bigotimes_{k=1}^n\bar{\sigma}_{i_k
j_k}^{(k)}\right),
\end{eqnarray}
where $\otimes_{k=1}^n A_k=A_1\otimes\cdots A_n$. $\rho^{*}$ is
the complex-conjugate of $\rho$. The $N_k$ dimensional matrices
$\sigma_{ij}^{(k)}$ and $\bar{\sigma}_{ij}^{(k)}$ act on the
$k^{th}$ subsystem with the entries defined as:
\begin{eqnarray*}
(\sigma_{ij}^{(k)})_{rs}&=&-\frac{i}{\sqrt{N_k-1}}(\delta_{ir}\delta_{js}-\delta_{jr}\delta_{is}),\\
(\bar{\sigma}_{ij}^{(k)})_{rs}&=&\frac{1}{\sqrt{N_k-1}}(\delta_{ir}\delta_{is}+\delta_{jr}\delta_{js}).
\end{eqnarray*}
\end{definition}

As a generalization of Jaeger's measure, we use the quadratic
function $tr[\rho F(\rho)]$ defined by the "flip" operation to
quantify the amount of entanglement in multipartite pure states.
One can verify that, owing to the factor $\frac{1}{\sqrt{N_k-1}}$
we introduce, this measure vanishes for arbitrary pure separable
states (see Theorem \ref{Vanish for pure seperable states}) of
general multipartite quantum systems, and, as a special case, is
reduced to Jaeger's Minkowskian norm entanglement measure for
multi-qubit states. However, as analyzed before, this measure will
become non-zero for mixed separable states that contain classical
correlation information\ucite{r16} among the subsystems, which is
related to the mixedness of the quantum states. We introduce the
following function:
$$M(\rho)=\left(\frac{2^n}{\prod_{k=1}^n N_k}-tr[\rho
\bar{F}(\rho)]\right),$$ to partially reflect the mixedness of the
states. Note that, except for the multi-qubit case, $M(\rho)$ is
different from the traditional mixedness expression
$M(\rho)=1-tr\rho^2$, because the measure is required to vanish
for pure separable states (see Theorem \ref{Vanish for pure
seperable states}) and fulfil some other conditions that we are
going to prove later (see Theorem \ref{Multipartite entanglement
measure}).

Finally, by subtracting $M(\rho)$ from the gross entanglement
measure $tr \rho F(\rho)$, we draw a quadratic quasi entanglement
measure
\begin{equation}\label{Quasi entanglement measure}
E_q(\rho)=\max\{ f(\rho),\,0 \},
\end{equation}
with the function
$f(\rho)$ defined as follows:
\begin{equation}\label{The density matrix type entanglement inequality function}
f(\rho)=tr[\rho F(\rho)]-\left(\frac{2^n}{\prod_{k=1}^{n}
N_k}-tr[\rho\bar{F}(\rho)]\right),
\end{equation}
or, in a more compact form:
\begin{equation}
f(\rho)=tr\rho [F(\rho)+\bar{F}(\rho)]-\frac{2^n}{\prod_{k=1}^{n}
N_k}.
\end{equation}

In the following parts of this paper, we will show that the
function $E_q(\rho)$ satisfies most of the conditions to be a
perfect entanglement measure, except that the function may be zero
for some entangled states. In this regard, we call the function
$E_q(\cdot)$ a {\it quasi entanglement measure}. For example, the
two-qubit Werner state\ucite{r20} \setlength{\mathindent}{0.5cm}
\begin{eqnarray*}
w=\frac{1}{4}I\otimes
I-\frac{1}{4}\cdot\frac{2\Phi+1}{3}\sum_{k=1}^3\sigma_k\otimes\sigma_k
\end{eqnarray*}
has been shown to be separable if and only if $-1\leq \Phi \leq
0$, but our measure gives a wider range
$\frac{-2-\sqrt{6}}{4}\approx-1.112\leq \Phi \leq
\frac{-2+\sqrt{6}}{4}\approx0.112$ for $E_q(w)=0$. Nevertheless,
it can be proven that $E_q(\rho)=0$, i.e. $f(\rho)\leq 0$, for all
separable states (see the proof of Theorem \ref{Multipartite
entanglement measure}). Therefore, the set of separable states is
a proper subset of the set of zero-entanglement states. The
relationship between these two sets is shown in Figure {\ref{Fig
of the function f}}.

\begin{figure}[h]
\centerline{ \includegraphics[width=3.0in,height=2.0in]{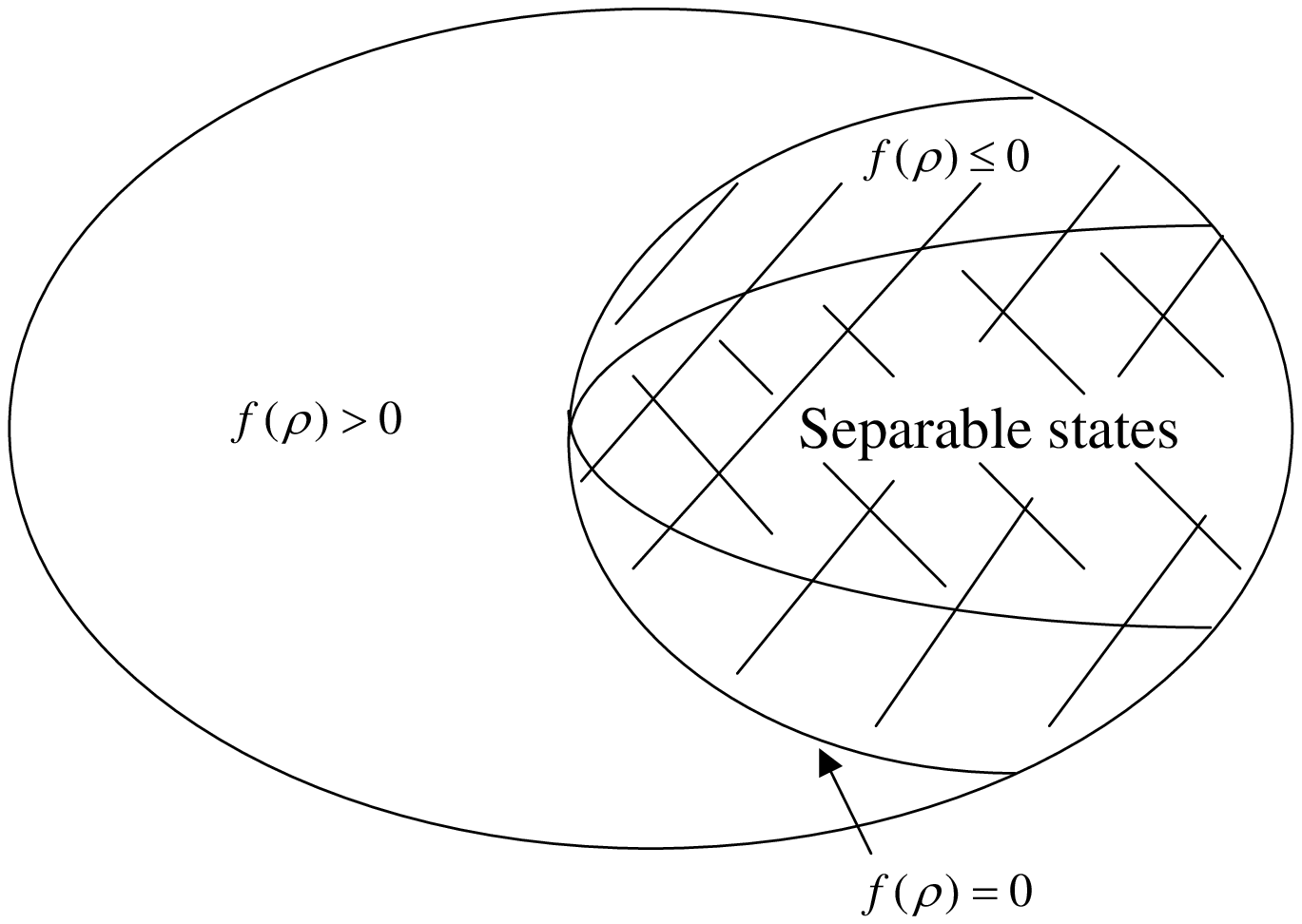}
} \caption{\scriptsize The relationship between the set $\{ \rho |
f(\rho)\leq 0 \}$ and the set of all separable states.}\label{Fig
of the function f}
\end{figure}

\begin{remark}
For states of which $f(\rho)$ are negative, the negative values
reflect the mixedness of the states due to the substraction of
$M(\rho)$. For example, for separable states, $f(\rho)$ is zero
for all pure separable states (see Theorem \ref{Vanish for pure
seperable states}), and goes below zero for mixed separable states
in which the minimum $\frac{2}{\prod_{k=1}^n N_k}(1-2^{n-1})$ is
reached at the completely mixed (separable) state
$\frac{1}{\prod_{k=1}^n N_k}I^{(1)}\otimes\cdots I^{(n)}$.
\end{remark}

To obtain a non-negative measure, we artificially cut off the
negative part of $f(\rho)$ in (\ref{Quasi entanglement measure})
so that all separable states have zero entanglement. Such a
definition amounts to a not so mathematically elegant non-smooth
function, which also appears in Wootters's concurrence
entanglement measure\ucite{r15}. Nevertheless, once we are able to
find a perfect $f(\rho)$ for which  $\rho$ satisfies $f(\rho)\leq
0$ if and only if $\rho$ is separable, $E(\rho)=\max\{f(\rho),0\}$
becomes a perfect entanglement measure. In this case, negative
$f(\rho)$ does not contain any entanglement information, because
this only happens to separable states. On the other hand, the
non-smoothness of such entanglement measures would help to explain
why entanglement may be lost in finite time\ucite{r5,r6,r7}.
Roughly speaking, consider a Markovian open system that
exponentially decays to an equilibrium separable distribution
$\rho_{\infty}$ such that $f(\rho_{\infty})<0$. Although the
equilibrium can be reached only in infinite time, the function
$f(\rho(t))$ will touch and go below zero within a finite time
interval because of the continuity of the function $f$. This is to
say, in this case, the entanglement in $\rho(t)$ will disappear
completely in finite time.

\section{Expanded coherence vector}
\label{s3}

In this section, we will represent the entanglement measure
$E_q(\rho)$ in the simpler coherence vector
picture\ucite{r27,r28,r29,r30,r31}, which has been widely applied
to describe the evolution of open quantum systems. For $N$-level
systems, the coherence vector of a density matrix $\rho$ is
derived as follows. Firstly, we choose an orthonormal basis $\{
\Omega_0, \Omega_1,\cdots,\Omega_{N^2-1} \}$ of $N\times N$
complex matrices with respect to the inner product $\langle
X,Y\rangle=tr(X^{\dagger}Y)$, where $\Omega_0=\frac{1}{\sqrt{N}}I$
is the normalized $N\times N$ identity matrix and
$\Omega_1,\cdots,\Omega_{N^2-1}$ are $N\times N$ normalized
traceless Hermitian matrices. A natural choice of the basis for
$N$-level systems is the generalization of Pauli matrices in
two-level systems:
$\{\Omega_{ij}^x,\,\Omega_{ij}^y,\,\Omega_{p}^z;1\leq i<j\leq N,\,
2\leq p\leq N\}$, where the entries of these matrices are:
\setlength{\mathindent}{1cm}
\begin{eqnarray}\label{The matrix basis of the N level systems}
(\Omega_{ij}^x)_{rs}=\frac{1}{\sqrt{2}}(\delta_{ir}\delta_{js}+\delta_{jr}\delta_{is}), \nonumber \\
(\Omega_{ij}^y)_{rs}=-\frac{i}{\sqrt{2}}(\delta_{ir}\delta_{js}-\delta_{jr}\delta_{is}),
 \\
(\Omega_p^z)_{rs}=\left\{\begin{array}{cc}
  \frac{1}{\sqrt{p(p-1)}}\delta_{rs} & r<p \\
  -\sqrt{\frac{p-1}{p}}\delta_{rs} & r=p \\
  0 & r>p \\
\end{array}\right. ,\nonumber
\end{eqnarray}
and $1\leq r,s \leq N$. Under this basis, any $N\times N$ complex
matrix $A$ can be expanded as
$A=\vec{a}\cdot\vec{\Omega}:=\sum_{i=0}^{N^2-1} a_i \Omega_i$
where $\vec{a}=(a_0,\cdots,a_{N^2-1})^T\in \mathbb{C}^{N^2}$
($\vec{a}\in \mathbb{R}^{N^2}$ if $A$ is Hermitian) and
$\vec{\Omega}=(\Omega_0,\cdots,\Omega_{N^2-1})^T$. Denote the
Hermitian density matrix $\rho$ as
$\rho=\bar{m}\cdot\vec{\Omega}$, where
$m_0=tr(\Omega_0\rho)\equiv\frac{1}{\sqrt{N}}$. The $N^2-1$
dimensional vector $m=(m_1,\cdots,m_{N^2-1})^T$ is called the
{\it{coherence vector}} of $\rho$, while
$\bar{m}=(m_0,m_1,\cdots,m_{N^2-1})^T$ is called the {\it{expanded
coherence vector}}. For convenience of the following computations
involving tensor product quantum states, we will frequently use
$\bar{m}$ instead of $m$. Obviously, since
$\{\Omega_i\}_{i=0,\cdots,N^2-1}$ is an orthonormal basis, we have
$\bar{m}^T\bar{m}=tr\rho^2\leq 1$, which implies that the expanded
coherence vector resides in the solid unit ball of
$\mathbb{R}^{N^2}$.

Generally, for a $n$-partite system of which the $k^{th}$
subsystem is $N_k$ dimensional, the orthonormal matrix basis can
be naturally chosen as the tensor product of basis matrices of the
subsystems:
$$\{ \Omega_{i_1}^{(1)}\otimes\cdots\otimes\Omega_{i_n}^{(n)}
;\,i_k=0,\cdots,N_k^2-1;k=1,\cdots,n\},$$ where $\{
\Omega_{i_k}^{(k)} \}_{i_k=0,\cdots,N_k^2-1}$ is the matrix basis
of the $k^{th}$ subsystem. Under this basis, a $n$-partite system
density matrix can be written as:
\begin{equation}\label{multipartite states in the coherence vector picture}
\rho=\sum_{\begin{array}{c}
  \scriptscriptstyle 0\leq i_1 \leq N_1^2-1 \\
  \scriptscriptstyle \cdots \\
  \scriptscriptstyle 0\leq i_n \leq N_n^2-1 \\
\end{array}}m_{i_1,\cdots,i_n}\Omega_{i_1}^{(1)}\otimes\cdots\Omega_{i_n}^{(n)},
\end{equation}
where the coefficients are:
\begin{equation}
m_{i_1\cdots i_n}=tr\left[\rho
\Omega_{i_1}^{(1)}\otimes\cdots\Omega_{i_n}^{(n)}\right].
\end{equation}

A novel property of the basis $\{
\Omega_{i_1}^{(1)}\otimes\cdots\Omega_{i_n}^{(n)} \}$ is that the
"flip" ("unflip") operations acting on the basis matrix
$\otimes_{k=1}^n \Omega_{r_k}^{(k)}$ can be decomposed into local
"flip" ("unflip") operations acting on the local basis matrix
$\Omega_{r_k}^{(k)}$, i.e.
 \setlength{\mathindent}{0.3cm}
\begin{eqnarray*}
&&F\left(\bigotimes_{k=1}^n \Omega_{r_k}^{(k)}\right)\\
&=&\sum_{\begin{array}{c}
  \scriptscriptstyle 1\leq i_1<j_1 \leq N_1 \\
  \scriptscriptstyle \cdots \\
  \scriptscriptstyle 1\leq i_n<j_n \leq N_n \\
\end{array}} \bigotimes_{k=1}^n \left( \sigma_{i_k
j_k}^{(k)} \Omega_{r_k}^{(k)}
\sigma_{i_k j_k}^{(k)} \right)\\
&=&\bigotimes_{k=1}^n \left( \sum_{1\leq i_k<j_k\leq N_k}
\left(\sigma_{i_k j_k}^{(k)}\right) \Omega_{r_k}^{(k)}
\left(\sigma_{i_k
j_k}^{(k)}\right) \right)\\
&=&\bigotimes_{k=1}^n F_k(\Omega_{r_k}^{(k)}),
\end{eqnarray*}
where $0\leq r_k\leq N_k^2-1$. For each local flip operation
$F_k$, one can examine by routine calculations that it keeps the
basis matrix $\Omega_0^{(k)}$ invariant, while flipping the other
traceless matrices $\Omega_{r_k}^{(k)}$ to
$-\frac{1}{N_k-1}\Omega_{r_k}^{(k)}$. In the coherence vector
picture, the local operation keeps $m^{(k)}_0$ invariant and
reverses the direction of the coherence vector $m^{(k)}$ with the
norm shrinking to its $\frac{1}{N_k-1}$ multiple. Similarly, the
local "unflip" operation $\bar{F}_k$ keeps both $m^{(k)}_0$ and
the direction of the coherence vector $m^{(k)}$ invariant, while
shortening the coherence vector to its $\frac{1}{N_k-1}$ multiple.

Therefore, the quadratic function (\ref{The density matrix type
entanglement inequality function}) can be expressed in the
coherence vector picture as follows:
\begin{equation}\label{The Coherence vector type entanglement inequality function}
f(\bar{m})=\bar{m}^T (S +\bar{S})\bar{m}-\frac{2^n}{\prod_{k=1}^n
N_k},
\end{equation}
where $S=S^{(1)}\otimes\cdots S^{(n)}$ and
$\bar{S}=\bar{S}^{(1)}\otimes\cdots \bar{S}^{(n)}$ with the
$N_k^2$ dimensional matrices:
\begin{equation}\label{Local flip and unflip operation in coherence vector picture}
S^{(k)}=\left(%
\begin{array}{cc}
  1 &  \\
   & -\frac{I_{N_k^2-1}}{N_k-1} \\
\end{array}%
\right),\quad \bar{S}^{(k)}=\left(%
\begin{array}{cc}
  1 &  \\
   & \frac{I_{N_k^2-1}}{N_k-1} \\
\end{array}%
\right)
\end{equation}
corresponding to local "flip" ("unflip") operations on local
expanded coherence vectors. Using this basic expression, one can
decompose the matrices $S$ and $\bar{S}$ into the following direct
sums: \setlength{\mathindent}{0.2cm}
\begin{eqnarray*}
  S &=& diag \left\{1,\bigoplus_{k;i_1,\cdots,i_k}(-1)^k \frac{I_{N_{i_1}^2-1}}{N_{i_1}-1}\otimes\cdots\otimes\frac{I_{N_{i_k}-1}}{N_{i_k}^2-1}\right\}, \\
  \bar{S} &=& diag \left\{1,\bigoplus_{k;i_1,\cdots,i_k}
  \frac{I_{N_{i_1}^2-1}}{N_{i_1}-1}\otimes\cdots\otimes\frac{I_{N_{i_k}^2-1}}{N_{i_k}-1}\right\},
\end{eqnarray*}
based on which we get the decomposition of $G=S+\bar{S}$:
\begin{eqnarray}\label{G in the coherence vector picture}
    G&=& 2\cdot diag \left\{1,\bigoplus_{2\nmid k;i_1,\cdots,i_k}
  \frac{0_{N_{i_1}^2-1}}{N_{i_1}-1}\otimes\cdots\otimes\frac{0_{N_{i_k}^2-1}}{N_{i_k}-1},\right. \nonumber \\
  && \left.\bigoplus_{2|k;i_1,\cdots,i_k}
  \frac{I_{N_{i_1}^2-1}}{N_{i_1}-1}\otimes\cdots\otimes\frac{I_{N_{i_k}^2-1}}{N_{i_k}-1}\right\}.
\end{eqnarray}
\setlength{\mathindent}{0.5cm}

Within the established coherence vector picture, we may obtain the
following equivalent expressions for future applications (see
Appendix \ref{Proof of the lemma: Basic notations in the expanded
coherence vector picture} for proof):

\begin{lemma}\label{Basic notations in the expanded coherence vector picture}
The fundamental concepts and operations in multipartite systems
can be rephrased as follows:
\begin{enumerate}
    \item[(1)] Separability: $\bar{m}$ corresponds to a separable state $\rho$ if and only if it can be written
    as:
$$\bar{m}=\sum_i p_i \bar{m}_i^{(1)}\otimes\cdots\otimes\bar{m}_i^{(n)},$$
where $\bar{m}_i^{(k)}$ is the $N_k^2$ dimensional expanded
coherence vector of a system density matrix $\rho_i^{(k)}$ of the
$k^{th}$ subsystem;
    \item[(2)] Local unitary operation can be represented by a tensor product matrix $\bar{O}^{(1)}\otimes\cdots \bar{O}^{(n)}$ acting on the expanded coherence vector $\bar{m}$, where $\bar{O}^{(k)}=diag(1,O^{(k)})$ and $O^{(k)}$ is a $N_k^2-1$ dimensional orthonormal matrix;
    \item[(3)] Local measurements can be expressed as the tensor product matrix $\bar{D}=\bar{D}^{(1)}\otimes\cdots
    \bar{D}^{(n)}$ acting on $\bar{m}$, where the matrix $\bar{D}^{(k)}$
    is $N_k^2$ dimensional. If the local
    measurements are POVM measurements, $\bar{D}^{(k)}=diag(1,D^{(k)})$ and the $N_k^2-1$ dimensional matrix
    $D^{(k)}$ is contractive, i.e. $D^{(k)T}D^{(k)}\leq I$.
\end{enumerate}
\end{lemma}

\section{Properties of the quasi entanglement measure}
\label{s4}

Firstly, we show that our quasi entanglement measure vanishes for
pure separable states:
\begin{theorem}\label{Vanish for pure seperable states}
For arbitrary pure separable states $\rho$, we have:
$$tr[\rho F(\rho)]=0,\quad \frac{2^n}{\prod_{k=1}^n N_k}-tr[\rho \bar{F}(\rho)]=0,$$
which means $E_q(\rho)=f(\rho)=0$.
\end{theorem}
\noindent{\bf Proof:} From the first item in lemma \ref{Basic
notations in the expanded coherence vector picture}, the coherence
vector of a pure separable state $\rho$ can be written as:
$$\bar{m}=\bar{m}^{(1)}\otimes\cdots\otimes\bar{m}^{(n)},$$
where $\bar{m}^{(k)}$ is the $N_k^2$ dimensional expanded
coherence vector of a pure state $\rho^{(k)}$ of the $k^{th}$
subsystem whose norm is 1, i.e.
$\bar{m}^{(k)T}\bar{m}^{(k)}=tr\rho^{(k)2}=1$.

Let
$\bar{m}^{(k)}=(m_0^{(k)},m^{(k)})^T=(\frac{1}{\sqrt{N_k}},m^{(k)})^T$,
we have
$$m^{(k)T} m^{(k)}=1-\frac{1}{N_k},$$
by which it can be deduced that:
\begin{eqnarray*}
\lefteqn{tr[\rho F(\rho)]=\bar{m}^T S
\bar{m}=\prod_{k=1}^{n}\bar{m}^{(k)T} S^{(k)} \bar{m}^{(k)}} \\
&& =\prod_{k=1}^{n}\left(
\frac{1}{N_k}-\frac{1}{N_k-1}m^{(k)T}m^{(k)} \right)=0,\\
\lefteqn{tr[\rho
\bar{F}(\rho)]=\bar{m}^T \bar{S}\bar{m}=\prod_{k=1}^{n}\bar{m}^{(k)T} \bar{S}^{(k)} \bar{m}^{(k)}} \\
&& =\prod_{k=1}^{n}\left(
\frac{1}{N_k}+\frac{1}{N_k-1}m^{(k)T}m^{(k)}
\right)=\frac{2^n}{\prod_{k=1}^n N_k}.
\end{eqnarray*}

\begin{theorem}\label{Multipartite entanglement
measure} The quadratic function $E_q(\rho)$ is a quasi
entanglement measure that possesses the following properties:
\begin{enumerate}
   \item [(1)] $E_q(\rho)\geq 0$; $E_q(\rho)=0$ if $\rho$ is
   separable;
   \item [(2)] Invariant under local unitary
   operations;
   \item [(3)] Non-increasing under local POVM
   measurements.
\end{enumerate}
\end{theorem}

\noindent{\bf Proof:} From $E_q(\rho)=\max\{f(\rho),0\}$ and lemma
\ref{Basic notations in the expanded coherence vector picture}, it
is sufficient to prove in the coherence vector picture:
\begin{itemize}
    \item $f(\bar{m})\leq 0$ if $\bar{m}=\sum_i p_i \bar{m}_i^{(1)}\otimes\cdots\otimes\bar{m}_i^{(n)}$;
    \item $f(\bar{O}\bar{m})=f(\bar{m})$ for arbitrary local unitary operation $\bar{O}=\bar{O}^{(1)}\otimes\cdots \otimes\bar{O}^{(n)}$, where $\bar{O}^{(k)}=diag(1,O^{(k)})$ and $O^{(k)}$ is an orthonormal matrix;
    \item $f(\bar{D}\bar{m})\leq f(\bar{m})$ for the operation $\bar{D}=\bar{D}^{(1)}\otimes\cdots \otimes\bar{D}^{(n)}$, where $\bar{D}^{(k)}=diag(1,D^{(k)})$ and the matrix $D^{(k)}$ is contractive, i.e. $D^{(k)T}D^{(k)}\leq I$.
\end{itemize}

For the first property, we can directly compute
that\setlength{\mathindent}{0.5cm}
\begin{eqnarray*}
&&f(\sum_i p_i \bar{m}_i^{(1)}\otimes\cdots
\otimes\bar{m}_i^{(n)})\\
&=&\sum_{i,j} p_i p_j\left(\prod_{k=1}^n \bar{m}_i^{(k)T}S^{(k)}
\bar{m}_j^{(k)} \right. +\\
&&\left.\prod_{k=1}^n \bar{m}_i^{(k)T} \bar{S}^{(k)}
\bar{m}_j^{(k)}\right)-\frac{2^n}{\prod_{k=1}^n N_k}.
\end{eqnarray*}
\setlength{\mathindent}{1.2cm}

From the fact that
$$\bar{m}_i^{(k)T}\bar{m}_i^{(k)}=\frac{1}{N_k}+m_i^{(k)T}m_i^{(k)}\leq
1 \Rightarrow \| m_i^{(k)} \|\leq \sqrt{1-\frac{1}{N_k}},$$ it can
be shown that $\bar{m}_i^{(k)T}S^{(k)}\bar{m}_j^{(k)}$ and
$\bar{m}_i^{(k)T}\bar{S}^{(k)}\bar{m}_j^{(k)}$ in the above
equation are both non-negative.

In fact, the first term is non-negative because
\setlength{\mathindent}{0.5cm}\begin{eqnarray*}
&&\bar{m}_i^{(k)T}S^{(k)}\bar{m}_j^{(k)}=\frac{1}{N_k}-\frac{m_i^{(k)T}m_j^{(k)}}{N_k-1} \\
&\geq& \frac{1}{N_k}-\frac{\| m_i^{(k)} \| \cdot \| m_j^{(k)}
\|}{N_k-1} \\
&\geq& \frac{1}{N_k}-\frac{1-N_k^{-1}}{N_k-1}=0,
\end{eqnarray*}
and the non-negativity of the latter term is because
\setlength{\mathindent}{0.5cm}\begin{eqnarray*}
 && \bar{m}_i^{(k)T}S^{(k)}\bar{m}_j^{(k)}=\frac{1}{N_k}+\frac{m_i^{(k)T}m_j^{(k)}}{N_k-1} \\
   &\geq& \frac{1}{N_k}-\frac{\| m_i^{(k)} \| \cdot \| m_j^{(k)} \|}{N_k-1} \geq 0.
\end{eqnarray*}

Employing the inequality
$$\prod_{k=1}^n a_k+\prod_{k=1}^n
b_k\leq\prod_{k=1}^n (a_k+b_k);\,\,\, a_k,b_k \geq 0,$$ we arrive
at the first property as follows:\setlength{\mathindent}{0.3cm}
\begin{eqnarray*}
&&f(\sum_i p_i \bar{m}_i^{(1)}\otimes\cdots \bar{m}_i^{(n)}) \\
&\leq& \sum_{i,j} p_i p_j\prod_{k=1}^n
\bar{m}_i^{(k)T}[S^{(k)}+\bar{S}^{(k)}]
\bar{m}_j^{(k)}-\frac{2^n}{\prod_{k=1}^n N_k} \\
&=&\sum_{i,j} p_i p_j\prod_{k=1}^n \frac{2}{N_k}
-\frac{2^n}{\prod_{k=1}^n N_k}=0.
\end{eqnarray*}\setlength{\mathindent}{0.5cm}

As to the second property, it is sufficient to prove:
$$
f(\bar{m})-f(\bar{O}\bar{m})=\bar{m}^T(G-\bar{O}^T G
\bar{O})\bar{m}=0,
$$
where $G=S+\bar{S}$. It can be deduced from $G-\bar{O}^T G
\bar{O}=0$ that can be easily verified from the commutativity
relationship:
$$[\bar{O}^{(1)}\otimes\cdots \otimes\bar{O}^{(n)},G]=0.$$

The third property requires that
\begin{eqnarray*}
f(\bar{m})-f(\bar{D}\bar{m})=\bar{m}^T(G-\bar{D}^T G
\bar{D})\bar{m}\geq 0,
\end{eqnarray*}
for which it is sufficient to prove that $G-\bar{D}^T G \bar{D}$
is a non-negative matrix.

From the expression of $G$ in (\ref{G in the coherence vector
picture}) and
\begin{eqnarray*}
\bar{D} & = & \bar{D}_1 \otimes \cdots \otimes\bar{D}_n \\
        & = & diag\left\{ 1,\bigoplus_{k;i_1,\cdots,i_k} D^{(i_1)}\otimes\cdots \otimes D^{(i_k)}
        \right\},
\end{eqnarray*}
 one can easily show that
$G-\bar{D}^T G \bar{D}$ is block-diagonal with the non-zero
diagonal blocks as:
$$\bigotimes_{l=1}^k \left[ \frac{1}{N_{i_l}-1}(I_{N_{i_l}^2-1}-D^{(i_l)T}D^{(i_l)}) \right],$$
where $2|k$. Thus $G-\bar{D}^T G \bar{D}\geq 0$ is obvious from
the contractive property of $D^{(i_k)}$. The end of proof.
$\quad\quad\Box$
\\[0.1cm]

Moreover, we can give an estimation of the bounds of the quadratic
quasi entanglement measure:
\begin{theorem}
For arbitrary quantum states $\rho$, the quasi entanglement
measure satisfies $0\leq E_q(\rho)\leq1$.
\end{theorem}
{\bf Proof:} The lower bound is obvious. Let $\bar{m}$ be the
corresponding expanded coherence vector of $\rho$. Suppose the
dimensions of the first $p$ subsystems $N_1,\cdots,N_p\geq 3$ and
$N_{p+1}=\cdots=N_n=2$ for the remaining subsystems. Writing
$\bar{m}^T(S+\bar{S})\bar{m}$ as the quadratic sum of the entries
of $\bar{m}$, i.e. $m_{i_1,\cdots,i_n}$ given in
(\ref{multipartite states in the coherence vector picture}), we
have:\setlength{\mathindent}{0.3cm}
\begin{eqnarray*}
&&\bar{m}^T(S+\bar{S})\bar{m} \\
&=&\frac{2}{\prod_{k=1}^n N_k}+2\sum_{\begin{array}{c}
  \scriptscriptstyle 1\leq s\leq n,2|s, \\
  \scriptscriptstyle 1\leq i_k\leq N_k^2-1 \\
\end{array}} \frac{m_{0\cdots i_1\cdots i_s\cdots 0}^2}{\prod_{k=1}^s
(N_{i_k}-1)}
\end{eqnarray*}

Divide the terms in the summation into three groups: the first
group only depends on the first $p$ subsystems whose dimensions
are no less than $3$, i.e. the non-zero indices $i_1\cdots i_s$ in
subscripts only come from the first $p$ subsystems; the second
group is related to both the first $p$ and the latter
two-dimensional subsystems, in which non-zero indices distribute
in both the two groups of subsystems with $i_1\cdots i_t$ from the
first and $i_{t+1}\cdots i_s$ from the second; the third group
depends only on the latter $n-p$ subsystems from which all
non-zero indices $i_1\cdots i_s$ come. Noting that $N_{i_k}-1$
will automatically disappear from the denominator for $i_k>p$
because $N_{i_k}=2$, we have:
\begin{eqnarray*}
&&\bar{m}^T(S+\bar{S})\bar{m} \\
&=&\frac{2}{\prod_{k=1}^n N_k}+2\sum_{\begin{array}{c}
  \scriptscriptstyle 1\leq s\leq n,2|s, \\
  \scriptscriptstyle 1\leq i_k\leq N_k^2-1 \\
\end{array}}\frac{m_{(0\cdots i_1\cdots
i_s\cdots 0)(0\cdots 0)}^2}{\prod_{k=1}^s{(N_{i_k}-1)}}\\
&&+2\sum_{\begin{array}{c}
  \scriptscriptstyle 1\leq s\leq n,2|s, \\
  \scriptscriptstyle 1\leq i_k\leq N_k^2-1 \\
\end{array}}\frac{m_{(0\cdot i_1\cdots
i_t\cdots 0)(0\cdots i_{t+1}\cdots i_s\cdots 0)}^2}{\prod_{k=1}^t{(N_{i_k}-1)}}\\
&&+2\sum_{\begin{array}{c}
  \scriptscriptstyle 1\leq s\leq n,2|s, \\
  \scriptscriptstyle 1\leq i_k\leq N_k^2-1 \\
\end{array}} m_{(0\cdots 0)(0\cdots i_1\cdots
i_s\cdots 0)}^2,
\end{eqnarray*}
Because, for the first two groups,
$$\frac{2}{\prod_{k=1}^s (N_{i_k}-1)}\leq 1, \frac{2}{\prod_{k=1}^t (N_{i_k}-1)}\leq 1,$$
we can derive that \setlength{\mathindent}{0.1cm}
\begin{eqnarray*}
&&\bar{m}^T(S+\bar{S})\bar{m}\\
&\leq&\frac{2}{\prod_{k=1}^n N_k}+\sum_{\begin{array}{c}
  \scriptscriptstyle 1\leq s\leq n,2|s, \\
  \scriptscriptstyle 1\leq i_k\leq N_k^2-1 \\
\end{array}}m_{(0\cdots i_1\cdots
i_s\cdots 0)(0\cdots 0)}^2\\
&&+\sum_{\begin{array}{c}
  \scriptscriptstyle 1\leq s\leq n,2|s, \\
  \scriptscriptstyle 1\leq i_k\leq N_k^2-1 \\
\end{array}} m_{(0\cdot i_1\cdots
i_t\cdots 0)(0\cdots i_{t+1}\cdots i_s\cdots 0)}^2
\end{eqnarray*}
\begin{eqnarray*}
&&+2\sum_{\begin{array}{c}
  \scriptscriptstyle 1\leq s\leq n,2|s, \\
  \scriptscriptstyle 1\leq i_k\leq N_k^2-1 \\
\end{array}} m_{(0\cdots 0)(0\cdots i_1\cdots
i_s\cdots 0)}^2 \\
&=&\frac{2}{\prod_{k=1}^n N_k}+\sum_{\begin{array}{c}
  \scriptscriptstyle 1\leq s\leq n,2|s, \\
  \scriptscriptstyle 1\leq i_k\leq N_k^2-1 \\
\end{array}}
(m_{0\cdot i_1\cdot i_s\cdot 0}^2 + m_{(0\cdot 0)(0\cdot i_1\cdot
i_s\cdot 0)}^2)\\
&\leq &\frac{2}{\prod_{k=1}^n N_k}+\sum_{\begin{array}{c}
  \scriptscriptstyle 1\leq s\leq n, \\
  \scriptscriptstyle 1\leq i_k\leq N_k^2-1 \\
\end{array}}(m_{0\cdot i_1\cdot i_s\cdot 0}^2 + m_{(0\cdot 0)(0\cdot i_1\cdot
i_s\cdot 0)}^2).
\end{eqnarray*}
From the equation
$$tr\rho^2=\bar{m}^T \bar{m}=\frac{1}{\prod_{k=1}^n N_k}+\sum_{\begin{array}{c}
  \scriptscriptstyle 1\leq s\leq n, \\
  \scriptscriptstyle 1\leq i_k\leq N_k^2-1 \\
\end{array}}
m_{0\cdots i_1\cdots i_s\cdots 0}^2,$$ it can be further
calculated that
\begin{eqnarray*}
&&\bar{m}^T(S+\bar{S})\bar{m}\\
&\leq& tr\rho^2+\frac{1}{\prod_{k=1}^n N_k}+\sum_{\begin{array}{c}
  \scriptscriptstyle 1\leq s\leq n, \\
  \scriptscriptstyle 1\leq i_k\leq N_k^2-1 \\
\end{array}}
m_{(0\cdots 0)(0\cdots i_1\cdots i_s\cdots 0)}^2 \\
&\leq& 1+\frac{1}{\prod_{k=1}^n N_k}+\sum_{\begin{array}{c}
  \scriptscriptstyle 1\leq s\leq n, \\
  \scriptscriptstyle 1\leq i_k\leq N_k^2-1 \\
\end{array}}
m_{(0\cdots 0)(0\cdots i_1\cdots i_s\cdots 0)}^2\\
&=& 1+\frac{1}{\prod_{k=1}^n N_k}+\sum_{\scriptscriptstyle 1\leq
i_k\leq N_k^2-1} m_{(0\cdots 0)(i_{p+1}\cdots i_n)}^2.
\end{eqnarray*}\setlength{\mathindent}{3cm}

Denote $\rho_{p+1,\cdots,n}$ the reduced density matrix for the
last $n-p$ subsystems. From (\ref{multipartite states in the
coherence vector picture}), one can show that:
\setlength{\mathindent}{0cm}
\begin{eqnarray*}
&&\rho_{p+1,\cdots,n}=tr_{1\cdots p}\rho \\
&=&\frac{1}{2^{n-p}} I+\sqrt{\prod_{k=1}^p N_k}
\sum_{i_{p+1}\cdot\cdot i_n} m_{(0\cdot 0)(i_{p+1}\cdot
i_n)}\Omega_{i_{p+1}}\otimes\cdots\Omega_{i_n},
\end{eqnarray*}
so we have
$$tr\rho_{p+1\cdots
n}^2=\frac{1}{2^{n-p}}+\left(\prod_{k=1}^p N_k\right) \sum_{1\leq
i_k \leq N_k^2-1}m_{(0\cdot 0)(i_{p+1}\cdot i_n)}^2\leq 1,$$ which
means \setlength{\mathindent}{0.3cm}
\begin{eqnarray*}
\sum_{1\leq i_k \leq N_k^2-1}m_{(0\cdots 0)(i_{p+1}\cdots i_n)}^2
\leq\frac{1}{\prod_{k=1}^p N_k}-\frac{1}{\prod_{k=1}^n N_k}.
\end{eqnarray*}\setlength{\mathindent}{3cm}
From this inequality, one obtains
that\setlength{\mathindent}{0.3cm}
\begin{eqnarray*}
&&\bar{m}^T (S+\bar{S})\bar{m} \\
&\leq& 1+\frac{1}{\prod_{k=1}^n
N_k}+\sum_{1\leq i_k \leq
N_k^2-1}m^2_{(0\cdots 0)(i_{p+1}\cdots i_n)} \\
&\leq & 1+\frac{1}{\prod_{k=1}^p N_k}\leq
1+\frac{2^p}{\prod_{k=1}^p N_k}\\
&=&1+\frac{2^n}{\prod_{k=1}^n N_k}.
\end{eqnarray*}
Therefore, $$0\leq
E_q(\rho)=\max\{f(\bar{m}),0\}\leq1+\frac{2^n}{\prod_{k=1}^n
N_k}-\frac{2^n}{\prod_{k=1}^n N_k}=1.$$ The end of proof.
$\qquad\qquad\qquad\qquad\qquad\qquad\qquad\qquad \Box$
\\[0.1cm]

Note that the upper bound can be reached for some multi-qubit
states. For example, the entanglement value of the well-known GHZ
state $\frac{1}{\sqrt{2^n}}\left( |0\cdots0>+|1\cdots1> \right)$
with even $n$ is equal to $1$.

\section{Examples}
\label{s5}

{\bf Example 1} Consider $n$-qubit quantum states that are widely
used in the quantum information theory. All the $n$ subsystems are
$2$-dimensional. It is not difficult to verify from (\ref{The
density matrix type entanglement inequality function}) that the
corresponding quadratic quasi entanglement measure can be
expressed as:
\begin{equation}
E_q(\rho)=\max\{ tr(\rho F(\rho))-[1-tr(\rho^2)],0 \},
\end{equation}
where $F(\rho)=\sigma_y^{\otimes n}\rho^*\sigma_y^{\otimes n}$ and
$\rho^*$ denotes the complex-conjugate of $\rho$. $\sigma_y$ is
the $y$-component of the well-known Pauli matrices.

For $n$-qubit pure states for which $tr\rho^2=1$, we have
$f(\rho)=tr(\rho F(\rho))$. As shown by Jaeger, $tr(\rho
F(\rho))\geq 0$ for physically meaningful states. Thus in this
case the quasi entanglement measure $E_q(\rho)=\max
\{f(\rho),0\}=f(\rho)$ is reduced to Jaeger's Minkowskian norm
entanglement measure $E(\rho)=tr(\rho F(\rho))$.
\\[0.1cm]

{\bf Example 2} Consider the entanglement measure of two-partite
systems with dimensions $N_1$ and $N_2$ respectively. The
two-partite "flip" operation $F(\rho)$ is equivalent to the
so-called universal state inverter\ucite{r33} as follows:
\begin{equation}\label{The universal state inverter}
F(\rho)=\frac{tr(\rho)I\otimes I-\rho_1\otimes
I-I\otimes\rho_2+\rho}{(N_1-1)(N_2-1)},
\end{equation}
where $\rho_1$ and $\rho_2$ denote the reduced density matrices of
the two subsystems.

In fact, in the coherence vector picture, we have:
\begin{eqnarray*}
\rho &=& \frac{1}{N_1 N_2}I+\sum_{1\leq i_1\leq N_1^2-1} m_{i_1
0}\,\,
\Omega_{i_1}\otimes\frac{1}{\sqrt{N_2}}I \\
&& +\sum_{1\leq i_2\leq
N_2^2-1} m_{0 i_2} \frac{1}{\sqrt{N_1}}I\otimes \Omega_{i_2} \\
&&+\sum_{1\leq i_k\leq N_k^2-1} m_{i_1
i_2}\,\,\Omega_{i_1}\otimes\Omega_{i_2},
\end{eqnarray*}\setlength{\mathindent}{0.2cm}
\begin{eqnarray*}
F(\rho) &=& \frac{1}{N_1 N_2}I-\frac{1}{N_1-1}\sum_{i_1} m_{i_1 0}
\,\,\Omega_{i_1}\otimes\frac{1}{\sqrt{N_2}}I \\
&& -\frac{1}{N_2-1}\sum_{i_2} m_{0 i_2} \frac{1}{\sqrt{N_1}}I\otimes \Omega_{i_2} \\
&&+\frac{1}{(N_1-1)(N_2-1)}\sum_{i_1,i_2} m_{i_1
i_2}\,\,\Omega_{i_1}\otimes\Omega_{i_2},
\end{eqnarray*}\setlength{\mathindent}{0.3cm}
and
\begin{eqnarray*}
\rho_1=\frac{1}{N_1}I+\sqrt{N_2}\sum_{1\leq i_1\leq N_1^2-1}m_{i_1
0}\,\, \Omega_{i_1}, \\
\rho_2=\frac{1}{N_2}I+\sqrt{N_1}\sum_{1\leq i_2\leq N_2^2-1}m_{0
i_2}\,\, \Omega_{i_2}.
\end{eqnarray*}
(\ref{The universal state inverter}) can be easily verified from
the above equations.

It can also be verified in the coherence vector picture that the
entanglement measure can be expressed in terms of the mixedness
function:
$$E_q(\rho)=\max\{\frac{2(N_1 M(\rho_1)+N_2 M(\rho_2)-N_1 N_2 M(\rho))}{(N_1-1)(N_2-1)N_1 N_2},0\},$$
where $M(\mu)=1-tr\mu^2$ is the mixedness function of some quantum
state $\mu$.

Suppose the two-partite state is a pure state, i.e. $M(\rho)=0$,
we find that the more entangled the global state is, the more
mixed the local states are. This shows that entanglement will
increase the uncertainties in local measurements.

\section{Conclusion}\label{s6}
In summary, we have developed a quadratic quasi entanglement
measure for the general multipartite quantum states. This measure
is a generalization of several well-known measures that have been
studied in the literature. The advantage of our measure is that it
can be expressed as a simple quadratic function of the coherence
vector that can be explicitly calculated. However, this measure is
still not perfect for most general quantum states, for which we
call it quasi entanglement measure, because it is not necessarily
non-zero for all entangled states and we are still not able to
prove the non-increasing property under more general LOCC
transformations except for local POVM measurements. Nevertheless,
the improvements of this measure comparing to the existing ones
open up many perspectives such as analysis of the mechanism of the
entanglement loss\ucite{r3,r4,r5,r6,r7}, and more importantly, the
control of preserving entanglement against environment-induced
decoherence effects, which used to be studied mainly from
numerical or experimental perspectives\ucite{r8,r9,r10}. These
remain to be studied in future work.
\\[0.2cm]

\begin{center}
\textbf{ACKNOWLEDGMENTS}
\end{center}

This research was supported in part by the National Natural
Science Foundation of China under Grant Number 60433050 and
60274025. T.J. Tarn would also like to acknowledge partial support
from the U.S. Army Research Office under Grant W911NF-04-1-0386.
\\[0.2cm]

\appendix
\section{Proof of the lemma \ref{Basic notations in the
expanded coherence vector picture}}\label{Proof of the lemma:
Basic notations in the expanded coherence vector picture}

(1) {\it Separable states in the expanded coherence vector
picture.}

By the definition of separable states, we
have:\setlength{\mathindent}{0.5cm}
\begin{eqnarray*}
\rho&=&\sum_i p_i
|\psi_i^1\rangle\langle\psi_i^1|\otimes\cdots\otimes|\psi_i^n\rangle\langle\psi_i^n|\\
&=&\sum_i p_i
\bar{m}_i^{(1)}\cdot\overrightarrow{\Omega}^{(1)}\otimes\cdots\bar{m}_i^{(n)}\cdot\overrightarrow{\Omega}^{(n)}\\
&=&(\sum_i p_i
\bar{m}_i^{(1)}\otimes\cdots\bar{m}_i^{(n)})\cdot\overrightarrow{\Omega}.
\end{eqnarray*}
Therefore, the expanded coherence vector of the separable states
must be in the form of $$\bar{m}=\sum_i p_i
\bar{m}_i^{(1)}\otimes\cdots\bar{m}_i^{(n)}.$$

(2) {\it Local unitary operation in the expanded coherence vector
picture.}

Firstly, the unitary operation of the $N$-level systems can be
expressed by the expanded coherence vector:
\begin{eqnarray*}
U\rho U^{\dagger} &=& U\left(
\frac{1}{N}I+\sum_{i=1}^{N^2-1}m_i\Omega_i\right)U^{\dagger}\\
&=&\frac{1}{N}I+\sum_{i=1}^{N^2-1}m_iU\Omega_i U^{\dagger}\\
&\equiv& \frac{1}{N}I+\sum_{i=1}^{N^2-1}\widetilde{m}_i\Omega_i.
\end{eqnarray*}
Obviously, $\widetilde{m}_0=m_0$. Denote
$m=(m_1,\cdots,m_{N^2-1})^T$,
$\widetilde{m}=(\widetilde{m}_1,\cdots,\widetilde{m}_{N^2-1})^T$,
we have
\begin{eqnarray*}
tr\rho^2=\frac{1}{N}+m^T m,\,\,\,tr(U\rho
U^{\dagger})^2=\frac{1}{N}+\widetilde{m}^T \widetilde{m},
\end{eqnarray*}
which imply that $\widetilde{m}^T \widetilde{m}=m^T m$. Therefore,
there exists an orthonormal matrix $O\in so(N^2-1)$, such that
$\widetilde{m}=Om$. Correspondingly, the expanded coherence
vectors satisfy that $\widetilde{\bar{m}}=\bar{O}\bar{m}$ with
$\bar{O}=diag(1,O)$.

Furthermore, it is easy to show that local unitary operations
$U_1\otimes\cdots U_n$ acting on the system density matrices can
be expressed as the tensor product operations
$\bar{O}=\bar{O}^{(1)}\otimes\cdots \bar{O}^{(n)}$ on the
corresponding expanded coherence vectors.

(3) {\it Local measurements in the expanded coherence vector
picture.}

For single partite case, measurements of $N$-level systems can be
expressed as linear trace-preserving Kraus maps:
$$\epsilon(\rho)=\sum_j L_j\rho
L_j^{\dagger}$$ with $\sum_j L_j^{\dagger}L_j= I$. One can always
express in the coherence vector picture the measurement by a
linear operation on the expanded coherence vector, i.e. the state
after a measurement can be written as
$\widetilde{\bar{m}}=\bar{D}\bar{m}$ where $\bar{D}$ is a constant
matrix with proper dimensions.

If the measurement is further restricted to be a POVM measurement,
i.e. $[L_j, L_j^{\dagger}]=0$, it can be calculated that
\setlength{\mathindent}{0.3cm}
\begin{eqnarray*}
\sum_j L_j\rho L_j^{\dagger} &=&\sum_j L_j
L_j^{\dagger}\frac{1}{N}I+\sum_j \sum_{i=1}^{N^2-1}m_iL_j\Omega_i
L_j^{\dagger}.
\end{eqnarray*}
\setlength{\mathindent}{0.5cm} The first term is equal to
$\frac{1}{N}I$ because $L_j L_j^{\dagger}=L_j^{\dagger}L_j$ and
$\sum_j L_j^{\dagger} L_j=I$. The second term can be expanded as a
linear combination of $\Omega_i,i=1,\cdots,N^2-1$, because it is
traceless. Thus, for a POVM measurement, we have
$$
\sum_j L_j\rho
L_j^{\dagger}=\frac{1}{N}I+\sum_{i=1}^{N^2-1}\widetilde{m}_i\Omega_i.
$$

Writing $\bar{D}$ as
$$\bar{D}=\left(%
\begin{array}{cc}
  a & h^T \\
  g & D \\
\end{array}%
\right),$$ one can show that
\begin{eqnarray*}
\bar{D}\bar{m}&=&(a m_0+h^T m, m_0 g+Dm)^T\\
&=&(\widetilde{m}_0,\widetilde{m}^T)^T
\end{eqnarray*}

The trace-preserving property requires that
$\widetilde{m}_0=m_0=\frac{1}{\sqrt{N}}$, which implies $a=1$ and
$h^T=0$. Further, let $m=0$, i.e. $\rho=\frac{1}{N}I$, it can be
verified that $\epsilon(\rho)=\frac{1}{N}I$. This is to say:
$$0=\widetilde{m}=Dm+m_0
g=m_0 g=\frac{1}{\sqrt{N}} g,$$ so we have $g=0$. Therefore,
$\bar{D}$ can be written in a block-diagonal form
$\bar{D}=diag(1,D)$.

The fact that $D$ is contractive for the POVM measurement
$\epsilon(\rho)$ can be proved by showing that
$$
tr\epsilon(\rho)^2\leq tr\rho^2.
$$

To prove this fact, note that the inequality
$tr(A-B)(A^{\dagger}-B^{\dagger})\geq 0$ gives:
$$trBA^{\dagger}+trAB^{\dagger}\leq
trAA^{\dagger}+trBB^{\dagger}.$$

Let $A=\rho L_j^{\dagger}L_k$ and $B=L_j^{\dagger}L_k\rho$.
Applying the above inequality together with the properties
$[L_i,L_i^{\dagger}]=0$ and $\sum_i L_i^{\dagger} L_i=I$, we have
\setlength{\mathindent}{0.2cm}
\begin{eqnarray*}
&&tr\epsilon(\rho)^2=\sum_{j,k} trL_j\rho L_j^{\dagger} L_k\rho L_k^{\dagger} \\
&=&\frac{1}{2}\sum_{j,k}\left( tr(\rho L_j^{\dagger}L_k \rho
L_k^{\dagger}L_j)+ tr(L_j^{\dagger}L_k \rho
L_k^{\dagger}L_j\rho)\right)\\
&\leq&\frac{1}{2}\sum_{j,k}\left(
tr(\rho L_j^{\dagger}L_k L_k^{\dagger}L_j\rho)+
tr(L_j^{\dagger}L_k \rho^2 L_k^{\dagger}L_j)\right)
\\
&=&\sum_{j,k} tr\rho^2(L_j^{\dagger}L_k
L_k^{\dagger}L_j)=tr\rho^2.
\end{eqnarray*}
\setlength{\mathindent}{0.5cm}

Thus it can be calculated that
\begin{eqnarray*}
&&\frac{1}{N}+m^T D^T D
m=\bar{m}^T\bar{D}^T\bar{D}\bar{m}=tr\epsilon(\rho)^2\\
&\leq& tr\rho^2=\bar{m}^T \bar{m}=\frac{1}{N}+m^T m,
\end{eqnarray*}
so we have $m^T D^T D m\leq m^T m$ for any $N^2-1$ dimensional
vector $m$ which means $D^T D\leq I$.

For $n$-partite case, a local POVM measurement can be written as:
$$\epsilon(\rho)=\sum_{i_1\cdots i_n} L_{i_1}^{(1)}\otimes\cdots L_{i_n}^{(n)}\rho L_{i_1}^{(1)\dagger}\otimes\cdots L_{i_n}^{(n)\dagger},$$
where $\sum_{i_k}L_{i_k}^{(k)\dagger}L_{i_k}^{(k)}=I$ and
$[L_{i_k}^{(k)},L_{i_k}^{(k)\dagger}]=0$. One can decompose
$\epsilon(\rho)$ into the product of the local operations,
$$\epsilon_k(\rho)=\sum_{i_k}M_{i_k}\rho M_{i_k}^{\dagger},$$
where $M_{i_k}=I\otimes\cdots L_{i_k}^{(k)}\otimes\cdots I$. It is
easy to verify that the resulting expanded coherence vector of
$\epsilon(\rho)$ can be written as
$\bar{D}\bar{m}=(\bar{D}^{(1)}\otimes\cdots
\bar{D}^{(n)})\bar{m}$, where $\bar{D}^{(k)}=diag(1,D^{(k)})$, and
$D^{(k)T} D^{(k)}\leq I$.
\\[0.1cm]

\end{document}